# Satellite Broadcasting Enabled Blockchain Protocol: A Preliminary Study

Ying-Hao Zhang
School of Computer Science and Engineering
Southeast University
Nanjing, China

Xiao Fan Liu
Department of Media and Communication
City University of Hong Kong
Hong Kong SAR, China
School of Computer Science and Engineering
Southeast University
Nanjing, China
e-mail: xf.liu@cityu.edu.hk

***Abstract*—Low throughput has been the biggest obstacle of large-scale blockchain applications. During the past few years, researchers have proposed various schemes to improve the systems' throughput. However, due to the inherent inefficiency and defects of the Internet, especially in data broadcasting tasks, these efforts all rendered unsatisfactory. In this paper, we propose a novel blockchain protocol which utilizes the satellite broadcasting network instead of the traditional Internet for data broadcasting and consensus tasks. An automatic resumption mechanism is also proposed to solve the unique communication problems of satellite broadcasting. Simulation results show that the proposed algorithm has a lower communication cost and can greatly improve the throughput of the blockchain system. Theoretical estimation of a satellite broadcasting enabled blockchain system's throughput is 6,000,000 TPS with a 20 gbps satellite bandwidth.**

***Keywords-blockchain; satellite broadcasting; throughput***

## I. INTRODUCTION

Blockchain technology has attracted much attention from academia and industry after the success of *Bitcoin*. Because of its special chained data structure, consensus algorithm and incentive mechanism, blockchain has some unique features, such as immutability, no need for absolute trust between nodes, etc. Although the blockchain system has some advantages that the traditional database does not have, low throughput has been the biggest obstacle to its large-scale application. *Bitcoin* produces a block in about 10 minutes with a throughput of about only 7 TPS (transactions per second) [1]. Meanwhile, today's centralized payment systems (such as *MasterCard* or *Visa*) can handle more than 24,000 TPS [2], which is three to four orders of magnitude higher than *Bitcoin*. *Ethereum*, widely considered as the second generation of blockchain, has a throughput about merely 15 TPS [3], which is still far lower than the centralized payment system.

Consensus algorithms of proposed blockchain systems must tolerate Byzantine errors [4]. It results in a higher concurrency cost and lower throughput compared to traditional database. Extensive research put efforts to optimize the consensus algorithm to improve blockchain's throughput. These results can be divided into four categories. 1) Modifying the underlying data structure to avoid forks. For example, *SPECTRE* [5] and *PHANTOM* [6] used directed acyclic graphs to replace the chain structure in the original design. Each block can point to multiple predecessors, making the entire ledger a directed acyclic graph. Then a linear sorting of all blocks is formed through the sorting algorithm, which avoids forks. 2) Weakening the decentralization requirement, that is, reducing the set of nodes participating in the leader election, such as *ByzCoin* [7] and *Algorand* [8]. The former chooses a group of nodes that have successfully mined within a certain period of time to form a committee. The members of the committee collectively decide to commit a block through the Byzantine consensus algorithm. The latter uses verifiable random functions to select committee members in a private and non-interactive way. 3) Separating leader election from record transactions and extend the duration of leadership to avoid frequent and relatively time-consuming leader elections. For example, *Bitcoin-ng* [9] divides blocks into two categories: key blocks and micro blocks. Key blocks contain only the proof of effort to obtain the leadership. The node's leadership status will last until the next key block is generated. Micro blocks only record transactions, which are verified by the leader and added to the blockchain. 4) Sharding, such as *OmniLedger* [10] and *Monoxide* [11]. The whole network is divided into several shards, each maintains a non-intersecting shard transaction record as a single-chain consensus system. In other words, there are multiple independent and parallel instances of single-chain consensus systems running concurrently in the whole network. Assuming that the entire network is divided into $K$ shards and there are no cross-shard transactions, the throughput of the entire network is $K$ times that of a single-chain consensus system.

However, these algorithms have either a limited improvement to throughput, with a maximum of about 900 TPS [8], or a visa-level throughput but lower security level [10,11].

The reason that previous efforts underachieved is that a large amount of data broadcasted is needed when keeping distributed systems consistent, and that the Internet has a poor efficiency when performing broadcasting tasks. Broadcasting in the Internet is not executed by multicast, i.e., group communication where data transmission is addressed to a group of destination computers simultaneously, but rather through replaying between single IP addresses. In other words, broadcasting is emulated by a group of unicasts. The internet also exposes its components to malicious attacks, especially Byzantine attacks, when emulating broadcasting by multiple unicasts [12]. In a blockchain system, the nodes do not have trust over each other. Therefore, consensus algorithms in blockchain not only need to maintain the data consistency of honest nodes, but also need to resist Byzantine attacks. The additional mechanisms require more resources and thus limits throughput.

Considering that the Internet is not an ideal communication channel for blockchain systems, this paper proposes a blockchain protocol utilizing satellite broadcasting network. In a satellite broadcasting network, all nodes can directly transmit messages to all other nodes through satellite broadcasting with only one hop of data transmission. There is no possibility of Byzantine attack, and the communication efficiency is greatly improved compared with the Internet.

Although satellite broadcasting has many advantages over Internet broadcasting, its unique physical constrain leads to some specific communication problems. For example, satellite broadcasting may be interrupted for a short period of time when a sun transit outage or an eclipse occurs. The unavailability of a single channel in traditional Internet only lead to the interruption of local network. While the interruption of satellite broadcasting may disable the communication between nodes and satellites in a large area, thus making the system temporarily unusable. Therefore, an automatic restart mechanism is also proposed in this paper for resuming data synchronization among nodes after extreme communication glitches happen.

This paper is organized as follows. Section II introduces the background of satellite communication. Section III introduces the consensus algorithm design and the automatic restart mechanism. Section IV tests the performance of the proposed protocol in a simulated environment. Section V discusses and concludes.

## II. SATELLITE COMMUNICATION

Satellite communication generally has two operation modes. One is point-to-point communication mode, that is, receiving data from the source host and forwarding it to the target host. The other is broadcasting mode, in which the satellite receives data and sends it to all hosts in a wide area. The latter mode is similar to UDP broadcasting on the Internet, but the data passes through only the communication satellite during the broadcasting process.

Satellite broadcasting communication system is generally composed of uplink stations, a communication satellite and receiving stations. As shown in Figure 1, the function of an uplink station is to send data to the broadcasting satellite. The satellite is the core of the whole system. Its main function is to process the received data signal then relay the processed signal to the ground receiving stations. The receiving stations receive signals transmitted by satellites and forward them to the user via other communication channels compared with other communication methods, satellite communication broadcasting has three advantages. 1) Wide coverage and long communication distance. One geosynchronous satellite can cover 42.4% of the earth's surface, with a maximum communication distance of more than 10,000 kilometers. 2) Broadcasting mode. Any geographical location within the satellite's coverage can receive a satellite signal. This feature can be used to realize multiple access communication. 3) Communication is not restricted by geographical conditions. Satellite signal is available at places which cannot be easily covered by traditional communication methods, such as mountains, oceans, etc.

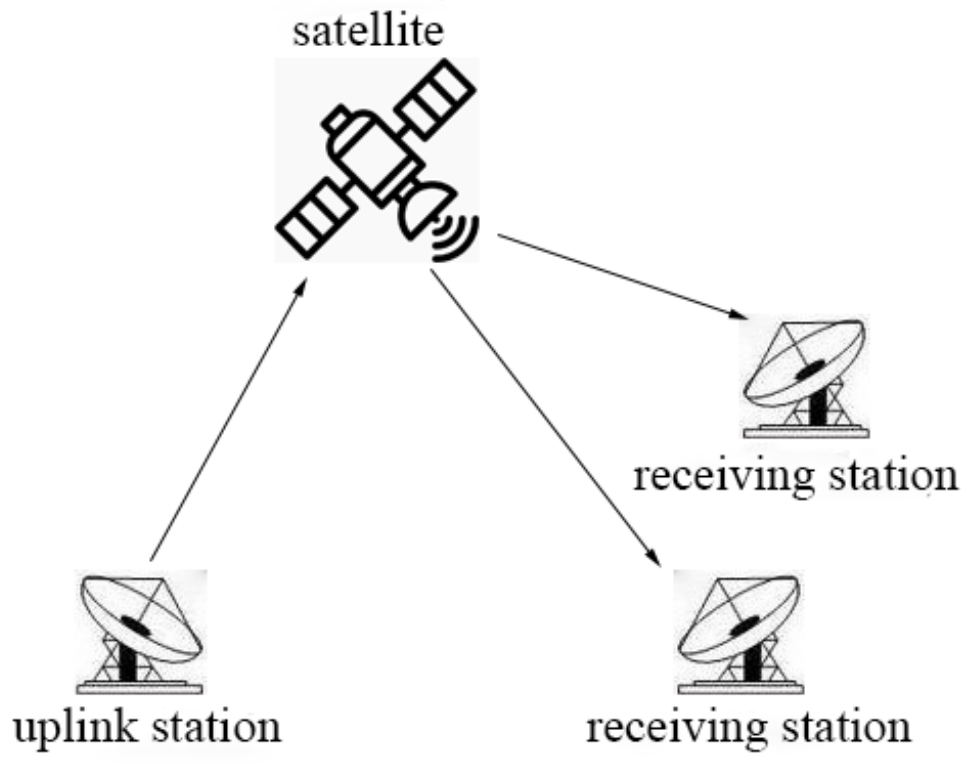

Figure 1. A satellite communication system.

However, satellite communication also has some disadvantages. 1) High construction cost. The manufacturing and launch of satellites have high costs. The construction of ground supporting facilities also need dedicated hardware. 2) Large transmission delay. The geostationary orbit satellite is 35860 km away from the ground [13]. Therefore, the round-trip transmission time is 239-278 ms [13].

This paper assumes a satellite broadcasting network which is structured as shown in Figure 2. The solid arrows represent the uplink channels from the system nodes to the satellite. The dotted arrows represent the satellite's broadcast channel. Apart from the communication channel shown in the figure, messages can also be sent and received between nodes through the Internet.

## III. BLOCKCHAIN PROTOCOL

The design of blockchain protocol considers several distinct features in a satellite broadcasting network. First, compared to the Internet, satellite broadcasting has the advantage that malicious nodes cannot use satellite broadcasting to launch Byzantine attacks. Therefore, the error model of the blockchain system can be simplified from Byzantine fault model to crash fault model, assuming that the uplink channels and communication channels between nodes are reliable. Second, the reachability and correctness of the broadcast channel are not guaranteed. Ground stations either accept the satellite signal or reject it, depending on error checking results of the received data. The problem of receiving error data can be transformed into packet loss problem. Third, the satellite broadcasts the data packets serially, so it can be considered that the packets received by all nodes are in the same order. The protocol is described in three parts: data broadcasting, data consensus and resumption from communication error.

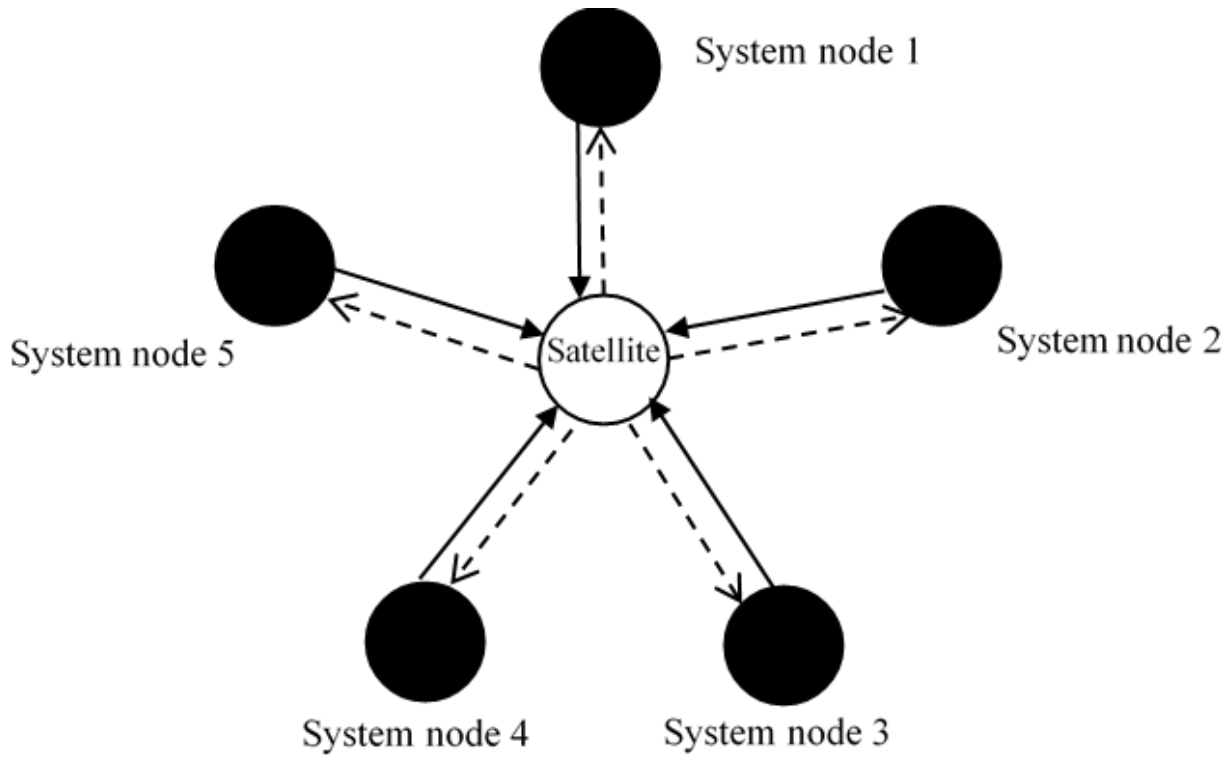

Figure 2. Schematic of satellite broadcasting network structure.

### A. Data Broadcasting

The system uses message as the smallest communication unit. Messages can contain different content as needed. For example, in a blockchain that supports smart contracts, it can be designed as a sequence of instructions containing several atomic operations; in a cryptocurrency, it can contain several transactions. Since ordinary user devices cannot interact with the satellite directly, clients must send data, i.e., as client messages, to one of the system nodes through the Internet. Then the system node will transmit the data, i.e., as data messages to the satellite for broadcasting. The processing flow when other ground stations receive a data message is shown in Figure 3. *M* is the block size, i.e., the number of messages in a block. The isSyncing flag is used to indicate whether the consensus round is underway or not. If the consensus round is currently in progress, the node will stop sending sync messages until the consensus process is finished, but the reception and processing of the client and data messages will continue.

Since satellite broadcasting messages can reach all nodes at the same time, there are only two periods for attack. One is in the process of the client sending data to the system node. The other is in the process that the system node sends the data to the satellite. Messages could be discarded or tampered by an attacker in either process. However, the attacker could not carry out a Byzantine attack because of satellite broadcasting. Tamper attacks can be prevented by encryption and/or digital digest. If a message is maliciously discarded, the client can select another system node and resend the message.

### B. Data Consensus

The satellite only sends signals to the ground but does not confirms the reception of them. Therefore, it is impossible to guarantee that all nodes can receive all signals, that is, a packet loss problem. In this section, a consensus mechanism is designed for making sure that data are synchronized across all nodes. The process of such a consensus round is shown in Figure 4.

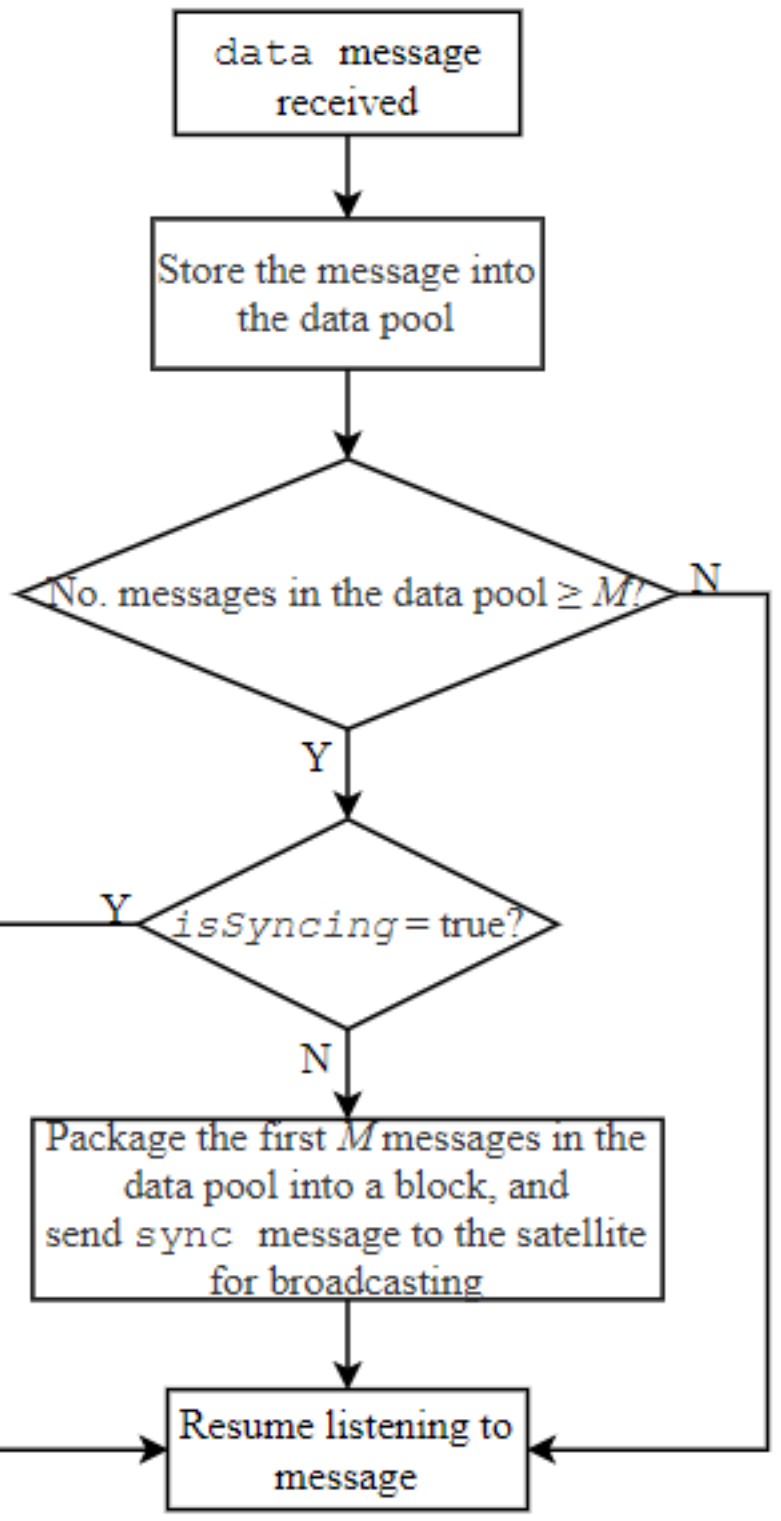

Figure 3. Data message processing flow

Each system node maintains a *nextIndex* variable that records the maximum block number that has been synchronized or the block number that is being synchronized, depending on whether the node is currently in the consensus round. At the beginning of each consensus round, system nodes filter sync messages that belong to the closed consensus round. sync messages contain block's index, block hash value, digital signature and other information necessary for reaching consensus. When the node is not in the consensus round and receives a sync message with an index greater than the node's *nextIndex*, it will start a new round of consensus. The node sets *nextIndex* to the index in the sync message. Then if the index in the next received sync messages is equal to *nextIndex* of the node, it means that the message belongs to that round of consensus, otherwise it does not.

The consensus process of each node is asynchronous. So, it may happen that a node receives a sync message with an index greater than *nextIndex* when performing a consensus round, i.e., indicating that a node had taken the lead to complete the current consensus round and started next round. In the case, the node will temporarily ignore the sync message until the current consensus round is complete. The purpose of this is to ensure that the data consensus is a linear process and avoid forks.

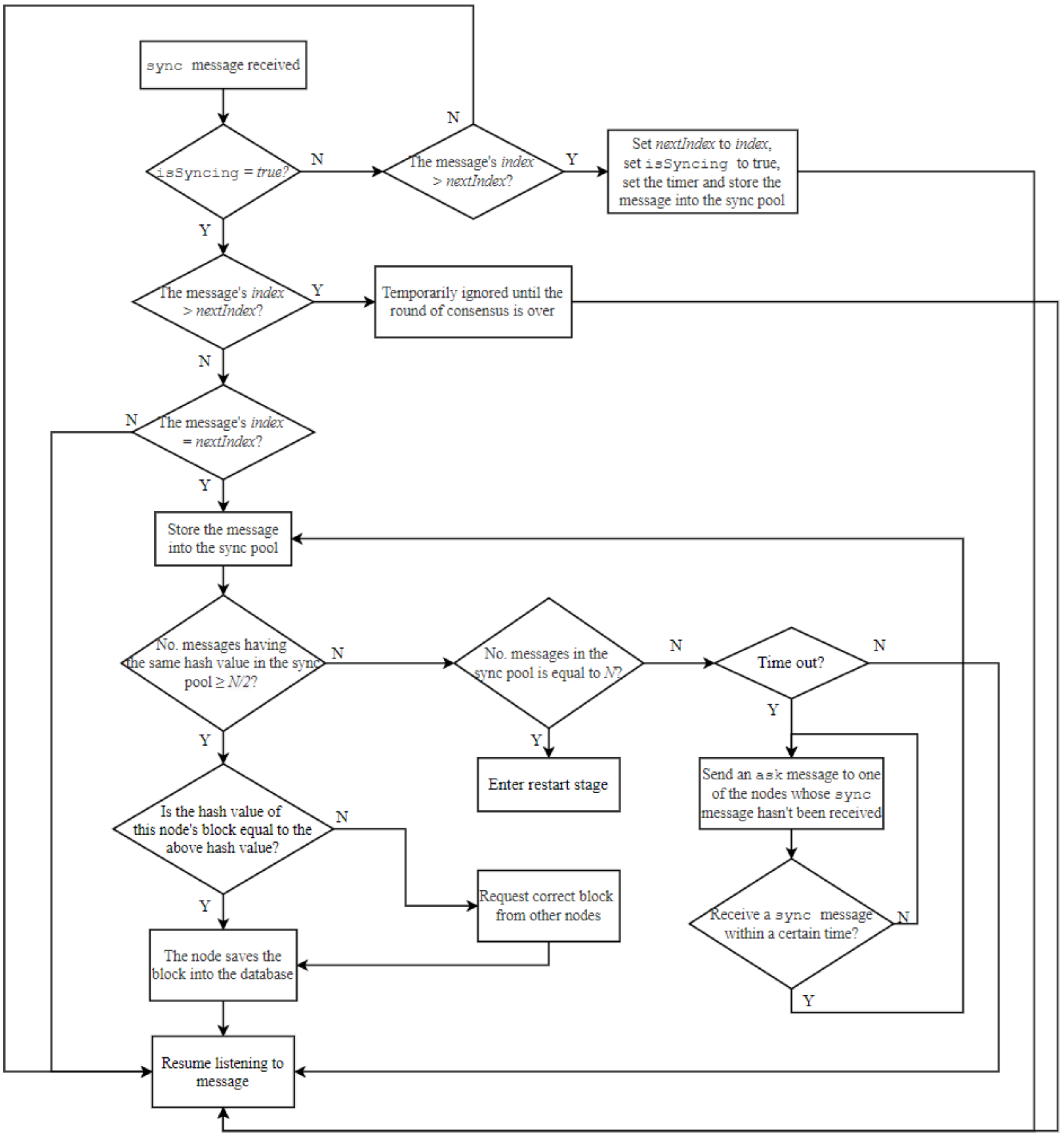

Figure 4. Flow chart of data consensus process

After filtering out invalid sync messages, system nodes will perform different operations depending on the sync message it receives. According to the number and content of sync messages received, there are three different situations, as shown in Figure 4.

*a)* Synchronization success: The node has received more than N/2 different sync messages and the hash values of those messages are the same, where N is the total numbers of nodes. It means that there is a block that is recognized by N/2 nodes. A block recognized by N/2 nodes is considered a correct block. If the hash value of the block is the same as the hash value of the block generated by the node, the node adds the block to the tail of its current blockchain. If not, the generated block is discarded and the correct block is requested from other nodes. At this point, in order to prevent the malicious node from replying to the wrong block to create data inconsistency, the node will calculate the hash value of the reply block and compare it with the correct hash value. If the two hash values are the same, the reply block is added to the tail of its blockchain. Otherwise the node selects another node and continues to request the correct block until the correct block is obtained. If the node has not yet produced a block, the node will continue the process of production. After the production finished, the node will check its correctness. That means the node is still allowed to finish the production. The reason to do this is that the sync message does not contain the entire block, but only the hash value of the block. If the block the current node produces is correct, the direct termination of the calculation will cause the node to request the block from other nodes, resulting in additional communication costs.

*b)* Synchronization failure: The node receives sync messages from all other nodes. But no more than half of sync messages have the same hash value. This indicates it is impossible to distinguish the correct block. Therefore, all nodes will abandon their block and enter the restart stage. This process will be introduced in details in Section III.

*c)* Synchronization uncertain: The node has not received sync messages from all other nodes within a certain period of time. And it is impossible to determine the correctness of blocks based on the information in the current sync pool. Therefore, this node will send an ask message though the Internet to one of the nodes which it hasn't received sync message from, and the target node will reply to the sync message of this round after receiving the ask message. When the node receives a reply, it will add the message in the sync pool. If the node does not receive a reply within a certain period of time, e.g., the target node may fail, an `ask` message will be sent to another node which it hasn't received sync message from. The node will repeat the process until it can make an accurate judgment.

### C. Automatic Resumption Mechanism

Special problems occur in satellite communication, due to its unique physical features. For example, when a sun transit outage or an eclipse occurs, satellite communication may enter a short-term interruption. The interruption of satellite communication may cause the failure of communication between nodes and satellites in a large area, making the system unable to reach consensus. Therefore, an automatic resumption mechanism is proposed. The mechanism can restore all nodes to the consensus state where the whole system can continue to run in case of the above problems. The mechanism flow is shown in Figure 5.

restart messages share the same broadcasting communication port with data messages. By utilizing the serialization of satellite broadcasting, we can ensure that the sequence of data messages received after the restart message is exactly the same in each node without packet loss. The system will restart from the last consensus round. All nodes that fail to reach consensus will send a restart message. The errorFlag flag makes sure that the nodes will only restart once after receiving the first restart message. It also helps to resist replay attacks and prevent malicious nodes from sending a restart message to enable the system to enter the restart process when the system is running normally. The drawback of this mechanism is that some messages are discarded and need to be resend by the clients.

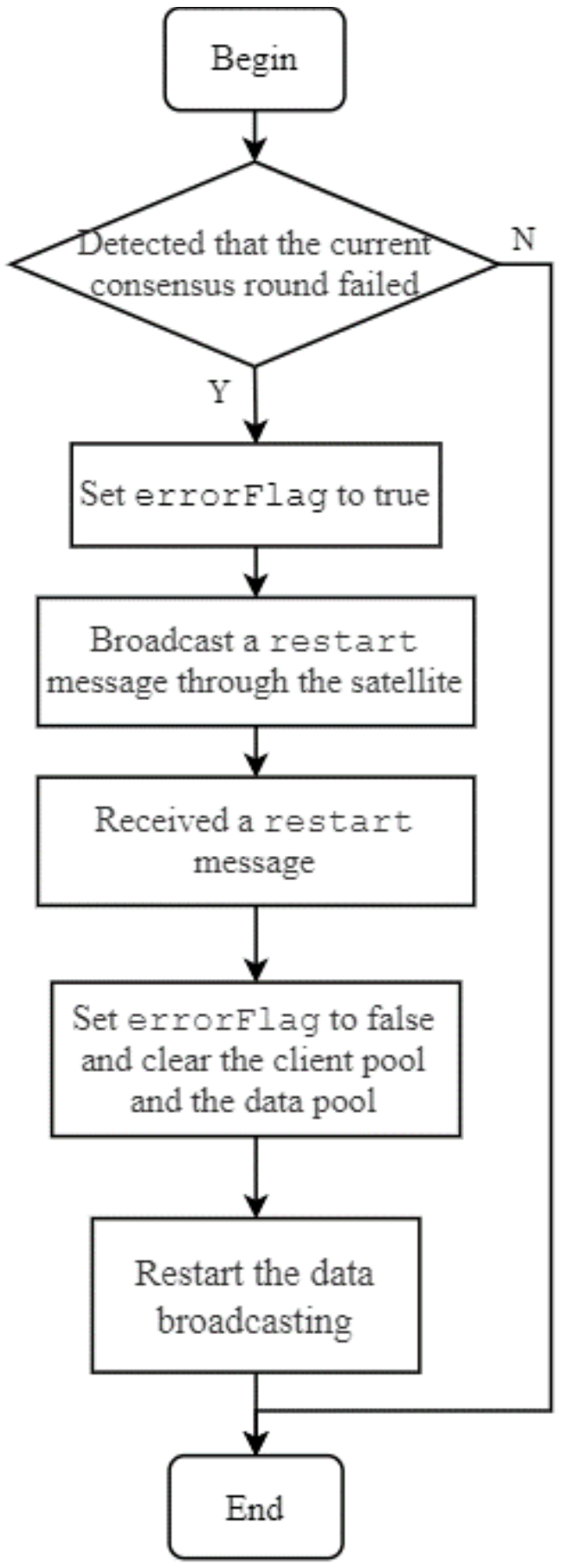

Figure 5. Automatic resumption mechanism flow

## IV. EXPERIMENTS

A LAN environment under a router imitates the satellite broadcasting network. In this environment, the broadcasted data can reach all the hosts in the LAN with only one hop through the router. This paper uses a LAN environment, as shown in Figure 6, to simulate the satellite broadcasting network. The black circles represent the system node, which are used to simulate receiving stations and a certain number of clients connected to them. The white circle represents the satellite node. The solid arrows represent the uplink channel of the receiving stations; the dotted arrows represent the broadcast channel of the satellite. Apart from the communication channel shown in the figure, messages can also be sent and received between nodes through the Internet. The broadcast messages are sent using UDP protocol, while others sent using TCP protocol.

Four machines were used in the experiment, three of which (IP address: 192.168.1.105, 192.168.1.175 and 192.168.1.189) ran system node programs and a certain number of client programs. One machine (IP address: 192.168.1.130) runs a simulated satellite program. A dedicated machine for satellite node is needed to simulate the serialization of the satellite broadcasting.

The client program continuously generates client messages and sends them to its local system node, i.e., server program. The server program runs the consensus algorithm to process all kinds of messages. The satellite program forwards the message received from the system node programs to the router for broadcasting.

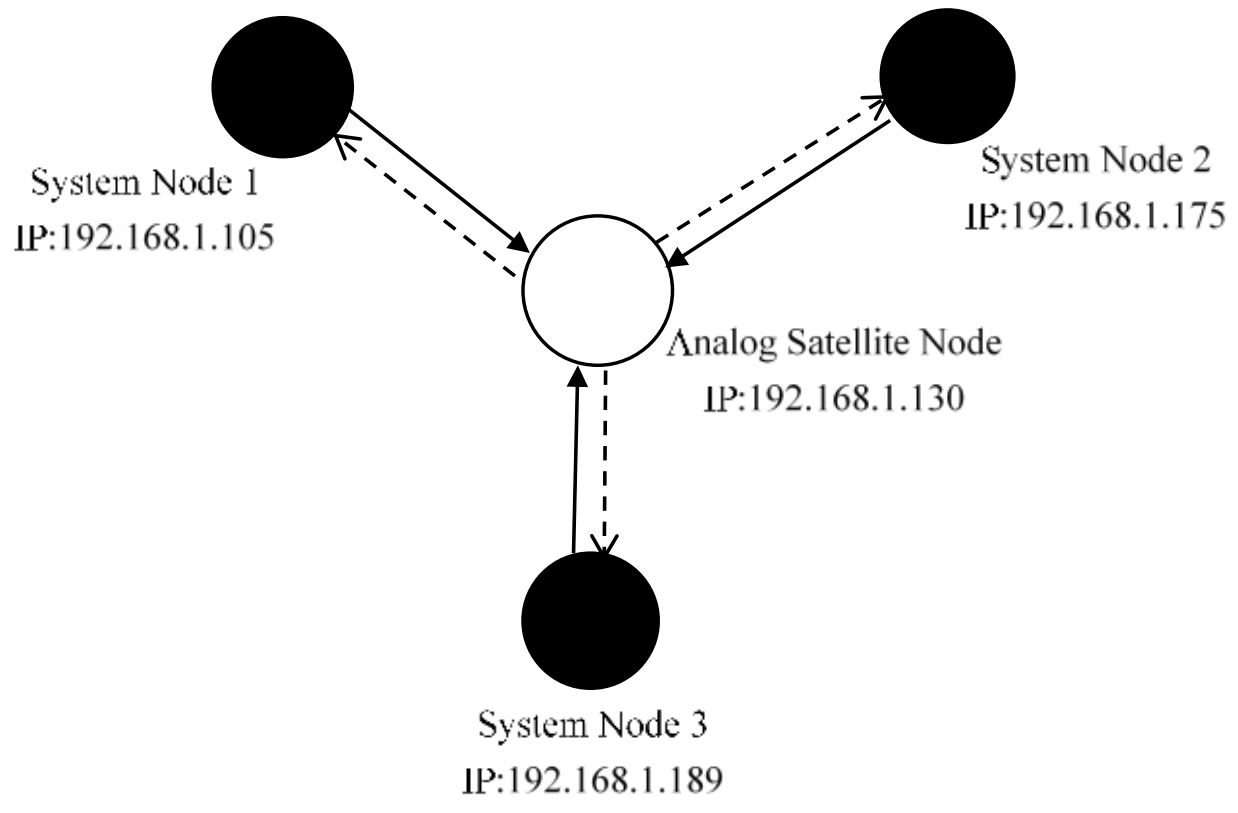

Figure 6. Analog experiment environment

## A. *Throughputt*

If a block contains *m* messages and the number of nodes is *n*, *m* data messages and *n* sync messages need to be broadcasted in order to synchronize a block. When *m* is much larger than *n* (in terms of total byte size), the communication bandwidth occupied by sync messages can be ignored. In summary, a data message needs to be broadcasted only once in order to synchronize itself.

Assuming that the satellite broadcast bandwidth is *B* and the receiving station uplink bandwidth is *U* (*U* < *B*), *msgSize* is the byte size of the data message, and the message synchronization rate, i.e., throughput, is *S*, then

$$\frac{U}{msgSize} \leq S \leq \frac{B}{msgSize} \quad (1)$$

If only one ground station is present in the blockchain system, the minimum throughput is constrained by its uplink bandwidth, i.e., $U/msgSize$. As the number of the ground stations increase, the total upstream rate of all ground stations will reach by not exceed *B*. At that moment, the throughput will be limited by the satellite broadcast bandwidth, meaning that the maximum of throughput is $B/msgSize$.

In the experiment setting, data generation rate is controlled by adjusting the number of client programs running concurrently. The size of the block is set to 10,000 messages. Throughput at different message generation rates is shown in the Figure 7. The experimental results verify the correctness of the above conclusions. As the rate of message generation increases, the system throughput remains approximately equal to the rate of message generation until the bandwidth bottleneck is reached. And the throughput is 10,167 TPS with 15 bytes transactions, when the bandwidth limit is about 1.2mbps.

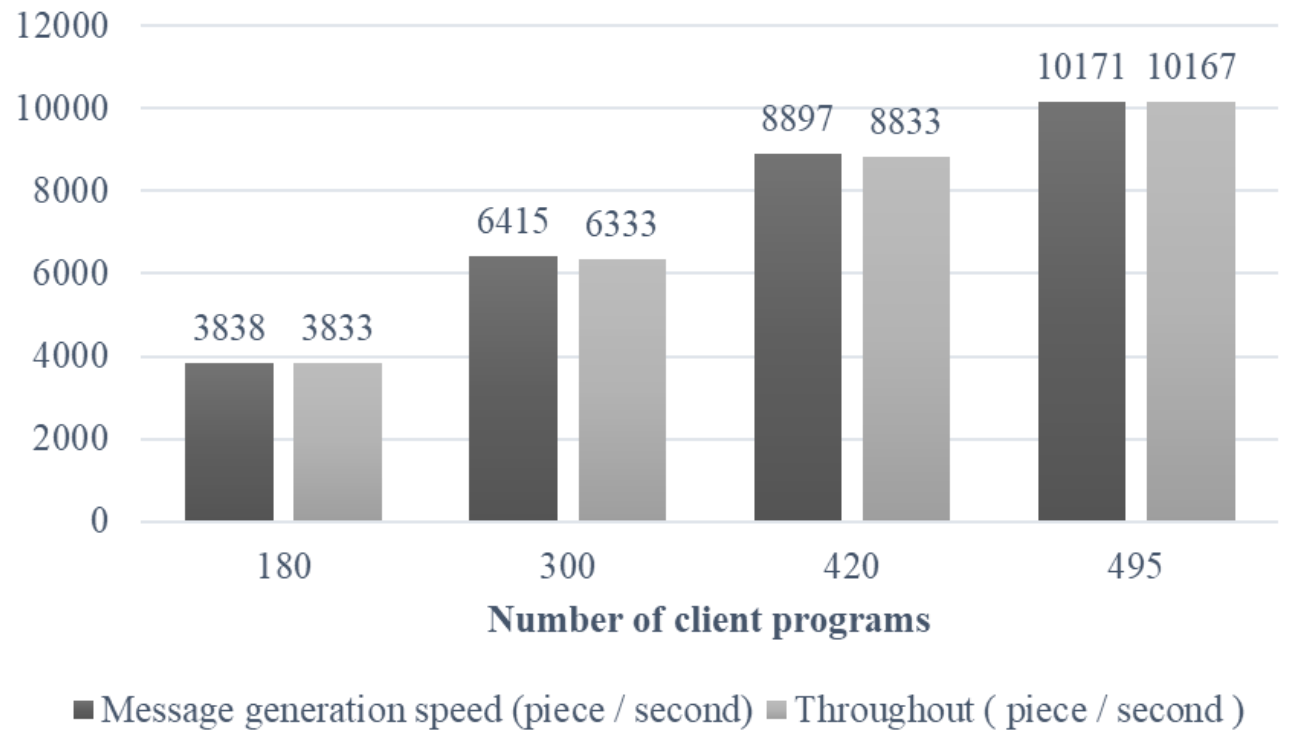

Figure 7. The results of throughput test

In practice, an *IPSTAR 1* Satellite has a total power supply of 14 kW and works on both Ku and Ka2 frequency bands. The satellite array antennae form 84 Ku-band cellular beams. The bidirectional transmission rate can reach 20Gbit/s. [13]. Assuming that the size of a data message is same as that in *Bitcoin* (about 300 bytes), a broadcast transmission rate of 20 Gbps could yield a theoretical throughput of about 8,000,000 TPS.

## B. *The Impact of Packet Loss Rate on Consensus*

Network packet loss is simulated by making nodes drop the data message it receives with a certain probability. The block size is set to 20 messages. The simulated packet loss rate was set as 3%, 6%, 9%, 15% and 30% respectively. A total of 10,000 consensus rounds were conducted each time. The ratio of success and failure times is shown in Figure 8. It is shown that when the packet loss rate is less than 10%, the loss of throughput is acceptable, which is about 43% at most. However, when the packet loss rate reaches 30%, the system is almost unusable.

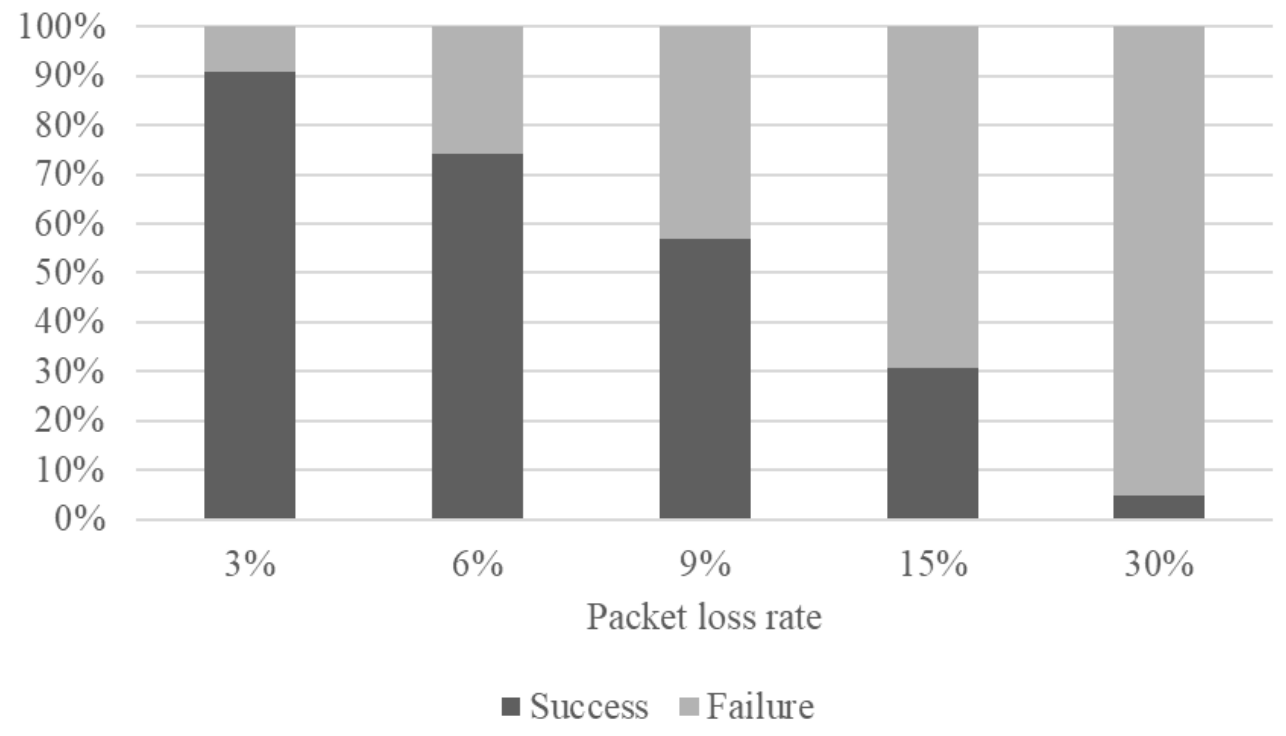

Figure 8. The impact of packet loss rate on consensus

In practice, the packet loss rate of China's Beidou satellite system under normal circumstances is about 4.38% [14]. As can be seen from Figure 8, when the packet loss rate is below 6%, the maximum loss of throughput is about 25%. Still using *IPSTAR 1* as an example, assuming that the packet loss rate is 6%, the theoretical throughput is about 6,000,000 TPS.

## V. Conclusion and Discussions

This paper proposed a blockchain protocol which utilizes satellite broadcasting as the blockchain system's communication channel. Satellite communication channel provides several unique features which are not present in the Internet. First, spatial locations of satellites are certain. Hence, communication time delay between ground stations and the satellite is certain. Second, ground stations are one and only one hop away from each other, with the satellite as the only relay. Third, the capacity, i.e., the number of ground stations, of a satellite broadcasting system is unlimited. It enables an unlimited scale of the blockchain system.

Considering the unique features of the satellite broadcasting system, the problem of Byzantine attack, which is caused by the IP replaying nature of the Internet, is not present in the satellite broadcasting system. Meanwhile, the serialization of satellite broadcasting can guarantee that the packets received by all nodes are in the same order. And by specifying that the node will discard packets that do not pass error checking, the problem of receiving error data can be transformed into packet loss problem. Therefore, the consensus algorithm proposed in this paper only has to deal with the packet loss problem, compared to Internet-based blockchain systems. This deterministic algorithm makes sure that the blockchain does not fork, therefore a relatively low confirmation delay. An automatic resumption mechanism was proposed to deal with the sun outage and eclipse problems in the satellite communication channel. This mechanism could resume the system to a normal state when these problems take place.

The nature of the satellite communication channel and our low-confirmation-delay data consensus algorithm resulted in a huge increase of system throughput compared to even the state-of-the-art Internet-based blockchain protocols, by magnitudes. The blockchain system can reach near-bandwidth throughput while preserving a fully de-centralized nature, given that the satellite remains neutral. A simulated satellite broadcasting channel with 20 Gbps bandwidth can yield about 6,000,000 TPS with 300 bytes transactions and a 6% packet drop rate.

A major limitation of this paper is that one broadcast satellite only covers less than half of the earth surface. Therefore, our proposed protocol is only suitable in a domestic or continental application. Nonetheless, it is superior in servicing as the infrastructure for high throughput requirement scenario, such as national electronic payments or other financial services systems.

A global satellite broadcasting system requires at least three geostationary satellites. The current inter-satellite routing technology in a multi-satellite system is similar to Internet routing, which inherits the defects of Internet routing. For example, when intersatellite links are not available, network partition occurs. These defects can cause additional consensus cost and safety issues. For example, if the algorithm runs in the permission-less blockchain, the system is vulnerable to a Sybil Attack [15]. Collaborations between blockchain protocol researchers and satellite manufactures should be called upon in the future to collectively solve these problems.

## Acknowledgement

Authors thank Prof. You-Ping Li, academician of the Chinese Academy of Engineering, for his valuable opinions in the preparation of this paper.